\documentclass[twocolumn,showpacs,amssymb,aps,prl,floatfix]{revtex4}

\usepackage{latexsym}
\usepackage{amsmath}
\usepackage{amsbsy}
\usepackage{epsfig}

\newcommand{\lapprox}{\raisebox{-0.5ex}{$\
\stackrel{\textstyle<}{\textstyle\sim}\ $}}
\newcommand{\gapprox}{\raisebox{-0.5ex}{$\
\stackrel{\textstyle>}{\textstyle\sim}\ $}}

\begin{document}

\title{A Quarkyonic Phase in Dense Two Color Matter?}

\author{
Simon Hands}

\affiliation{
       Department of Physics, Swansea University,
       Singleton Park, Swansea SA2 8PP, U.K.
       }

\author{Seyong Kim}
\affiliation{
       Department of Physics, Sejong University,
       Seoul 143-747, Korea. 
       }

\author{Jon-Ivar Skullerud}

\affiliation{
Department of Mathematical Physics, National University of Ireland Maynooth,
Maynooth, County Kildare, Ireland.}


\begin{abstract}

We present results from simulations of Two Color QCD with two Wilson quark
flavors in the presence of a quark chemical potential $\mu$ at two different
lattice spacings.
The
equation of state, conformal anomaly, superfluid order parameter and Polyakov
line are all discussed. Our results suggest that the transition from hadronic to quark
matter, and that from confined to deconfined matter occur at distinct values
of $\mu$, consistent with the existence of a quarkyonic phase in this model.

\end{abstract}

         
\maketitle

A quantitative understanding of cold dense baryonic matter,
ie. the region of the $(T,\mu)$ plane 
conventionally placed at the
lower right of the QCD phase diagram (here $\mu$ is the quark chemical
potential), 
remains an
outstanding problem in theoretical physics. 
There has been much discussion of possible exotic 
color superconducting (CSC) phases in this region, where
color-carrying degrees of freedom such as quarks or gluons are all gapped via a
Higgs-Meissner mechanism, and which may
also be superfluid or even crystalline
\cite{Alford:2007xm}.
This picture has its firmest
theoretical support at weak gauge coupling, where superconductivity is
understood as being due to the BCS condensation of diquark Cooper pairs. More
recently, an alternative scenario based on large-$N_c$ arguments has emerged, 
whose ground state at sufficiently high density has a restored chiral symmetry, 
so that thermodynamic properties are well-described by a degenerate
relativistic Fermi sea of quarks characterised by a Fermi momentum $k_F$, but which
remains in a confined ``quarkyonic'' phase so that all excitations necessarily carry
color-singlet quantum numbers 
\cite{McLerran:2007qj}.

It remains unclear whether the CSC and quarkyonic pictures are
truly distinct, or are in some sense complementary. However, it is salutary to
recall that  even
basic questions involving dense matter, such as the maximum stable mass of a neutron star, 
need
quantitative input about the equation of state of ultradense matter regardless
of the nature of the ground state; this
requires a controlled non-perturbative calculation.

The most reliable source of such information, lattice QCD, is in general
inoperable in this regime for the following reason. In Euclidean metric the 
QCD Lagrangian density for quarks reads
\begin{equation}
{\cal L}_{QCD}=\bar\psi(D\!\!\!\!/\,[A]+\mu\gamma_0+m)\psi\equiv
\bar\psi M\psi.
\end{equation}
It is straightforward to show $\gamma_5M(\mu)\gamma_5\equiv M^\dagger(-\mu)$,
implying $\mbox{det}M(\mu)=(\mbox{det}M(-\mu))^*$, and therefore that the path
integral measure is not real and positive for $\mu\not=0$.
Monte Carlo importance sampling, the mainstay of numerical lattice
QCD, is ineffective. It is helpful to consider what goes wrong if the real positive 
measure factor $\mbox{det}M^\dagger M$ as implemented in, eg. the 
hybrid Monte Carlo (HMC)
algorithm, is used. In QCD, while $M$ describes a color
triplet of quark fields $q$, the $M^\dagger$ factor describes color antitriplet
``conjugate quarks'' $q^c$. Gauge singlet bound states of the form $qq^c$ 
resemble
mesons, but carry non-zero baryon charge $B>0$.
The lightest such state is degenerate with the pseudo-Goldstone $\pi$-meson;
hence HMC simulations with $\mu\not=0$ predict an unphysical ``onset'' 
transition from the vacuum to a state with quark density $n_q>0$ at
$\mu_o\simeq{1\over2}m_\pi$. The resulting ground state is a Bose-Einstein
condensate (BEC) of diquark baryons, bearing no resemblance to nuclear
matter, which phenomenologically we know forms at $\mu_o\approx{1\over
N_c}m_{nucleon}$. The physical transition can only be found if the correct
complex path integral measure $\mbox{det}^2M$ is used, and must
result from extremely non-trivial cancellations between configurations with
differing phases  -- this has come to be known as the {\em Silver Blaze\/}
problem~\cite{Cohen:2003kd}.

In this paper we consider an alternative strongly-interacting theory, Two
Color QCD (QC$_2$D), in which the gauge group is SU(2).
Since $q$ and $\bar q$ live in
equivalent representations of SU(2), it follows that 
$\mbox{det}M(\mu)\equiv\mbox{det}\tau_2M^*(\mu)\tau_2$ 
is real and therefore the theory has a positive measure for an even number $N_f$
of 
quark flavors~\cite{Hands:2000ei}. 
Physically this is expressed through 
both $q\bar q$ mesons and $qq$, $\bar q\bar q$ baryons falling in the same
hadron mulitplets.
For sufficiently light quarks the scale hierarchy $m_\pi\ll m_\rho$
permits the use of chiral perturbation theory ($\chi$PT) in studying the
response of the lightest multiplet to $\mu\not=0$~\cite{Kogut:2000ek}. 
The key result
is
that for $\mu\geq\mu_o\equiv{1\over2}m_\pi$ a non-zero baryon charge density
$n_q>0$ does develop, 
along with a gauge-invariant superfluid order parameter which 
for $N_f=2$ reads $\langle qq\rangle\sim
\langle \psi^{tr}C\gamma_5\tau_2\epsilon_{ab}\psi\rangle\not=0$, where $\tau_2$
acts on color indices and $\epsilon_{ab}=-\epsilon_{ba}$ on flavor.
The resulting BEC is composed of weakly interacting $qq$
baryons with $J^P=0^+$.

For $\mu\geq\mu_o$ leading-order $\chi$PT predicts a smooth rotation of the chiral condensate
$\langle\bar\psi\psi\rangle$ into the superfluid condensate $\langle qq\rangle$
as $\mu$ increases~\cite{Kogut:2000ek}. In addition there is a quantitative prediction for quark
density:
\begin{equation}
n_q(\mu)=8N_f f_\pi^2\mu\left(1-{\mu_o^4\over\mu^4}\right),
\label{eq:xPT1}
\end{equation}
where the parameters $\mu_o$ and $f_\pi$ suffice to specify $\chi$PT at this
order.
It is possible to 
develop the thermodynamics of the system at $T=0$ more
fully, to extract pressure and energy density
\cite{Hands:2006ve}:
\begin{eqnarray}
p_{\chi PT}=\int_{\mu_o}^\mu n_q d\mu &=&
4N_f f_\pi^2\left(\mu^2+{\mu_o^4\over\mu^2}-2\mu_o^2\right);\label{eq:xPT2}\\
\varepsilon_{\chi PT}=-p+\mu n_q &=&
4N_f f_\pi^2\left(\mu^2-3{\mu_o^4\over\mu^2}+2\mu_o^2\right);\label{eq:xPT3}\\
(T_{\mu\mu})_{\chi PT}=\varepsilon-3p &=&
8N_f f_\pi^2\left(-\mu^2-3{\mu_o^4\over\mu^2}+4\mu_o^2\right)\!\!.\label{eq:xPT4}
\end{eqnarray}
Note that the trace of the stress-energy tensor $(T_{\mu\mu})_{\chi PT}<0$ 
for $\mu>\surd3\mu_o$.

These model results should be contrasted with those of another paradigm for cold dense
matter, namely a degenerate system of weakly interacting (thus presumably deconfined)
quarks populating a Fermi sphere up to some maximum momentum $k_F\approx
E_F=\mu$:
\begin{equation}
n_{SB}={{N_fN_c}\over{3\pi^2}}\mu^3;\;
\varepsilon_{SB}=3p_{SB}={{N_fN_c}\over{4\pi^2}}\mu^4.
\label{eq:SB1}
\end{equation}
In this picture superfluidity arises from condensation of diquark Cooper pairs
from within a layer of thickness $\Delta$ centred on the Fermi surface; hence
\begin{equation}
\langle qq\rangle\propto\Delta\mu^2.
\label{eq:BCS}
\end{equation}
Since $p_{SB}$ eventually exceeds $p_{\chi PT}$ as $\mu$ increases, the
degenerate system must be the more thermodynamically stable at high density.
In fact, since both $n_q$ and $\varepsilon$ are discontinuous at the point
where $p_{SB}=p_{\chi PT}$, 
this naive treatment predicts
the resulting deconfining transition is first order~\cite{Hands:2006ve}.

These considerations have motivated us to pursue lattice simulations of
QC$_2$D beyond the BEC regime, using $N_f=2$ flavors of Wilson
fermion. The quark action is
\begin{equation}
S=\sum_{i=1,2}\bar\psi_iM\psi_i+\kappa
j[\psi_2^{tr}(C\gamma_5)\tau_2\psi_1-h.c.],
\label{eq:Slatt}
\end{equation}
with
\begin{eqnarray}
M_{xy}=\delta_{xy}-\kappa&\sum_\nu&\Bigl[(1-\gamma_\nu)e^{\mu\delta_{\nu0}}U_\nu(x)\delta_{y,x+\hat\nu}\nonumber\\
&+&(1+\gamma_\nu)e^{-\mu\delta_{\nu0}}U^\dagger_\nu(y)\delta_{y,x-\hat\nu}\Bigr].
\end{eqnarray}
A conventional Wilson action was used for the glue fields. Further details 
can be found in \cite{Hands:2006ve}. 

Since Wilson fermions do not have a manifest chiral symmetry, we have
little to say at this stage about this aspect of the physics, 
which at high quark density should 
be of secondary
importance for phenomena near the Fermi surface; they do however carry a conserved
baryon charge due to the U(1)$_B$ symmetry $\psi\mapsto e^{i\alpha}\psi$,
$\bar\psi\mapsto\bar\psi e^{-i\alpha}$. 
Our initial runs on a $8^3\times16$ lattice with $\beta=1.7$, $\kappa=0.178$
corresponding to lattice spacing 
$a=0.230(5)$fm, $m_\pi a=0.79(1)$ and $m_\pi/m_\rho=0.779(4)$ have been described in
~\cite{Hands:2006ve}.
In this Letter we present data from runs on an
approximately matched $12^3\times24$ lattice with $\beta=1.9$, $\kappa=0.168$
corresponding to $a=0.186(8)$fm, $m_\pi a=0.68(1)$
and $m_\pi/m_\rho=0.80(1)$. The physical scale is set by equating the observed string
tension at $\mu=0$ to (440MeV)$^2$. Note that the physical temperature $T$ is
approximately 54(1)MeV for the smaller lattice, and 44(2)MeV for the larger.
We used a standard HMC
algorithm -- the only novelty is the inclusion of a
diquark source term (proportional to $j$ in Eqn.(\ref{eq:Slatt}));
this mitigates the impact of 
IR fluctuations in the superfluid regime and also enables the 
algorithm to change the sign of $\mbox{det}M$ for a single flavor, thus
maintaining ergodicity. All results presented here were obtained with $ja=0.04$;
ultimately the physical limit $j\to0$ must be taken.

\begin{figure}[ht]
    \centering
    \includegraphics[width=8.0cm]{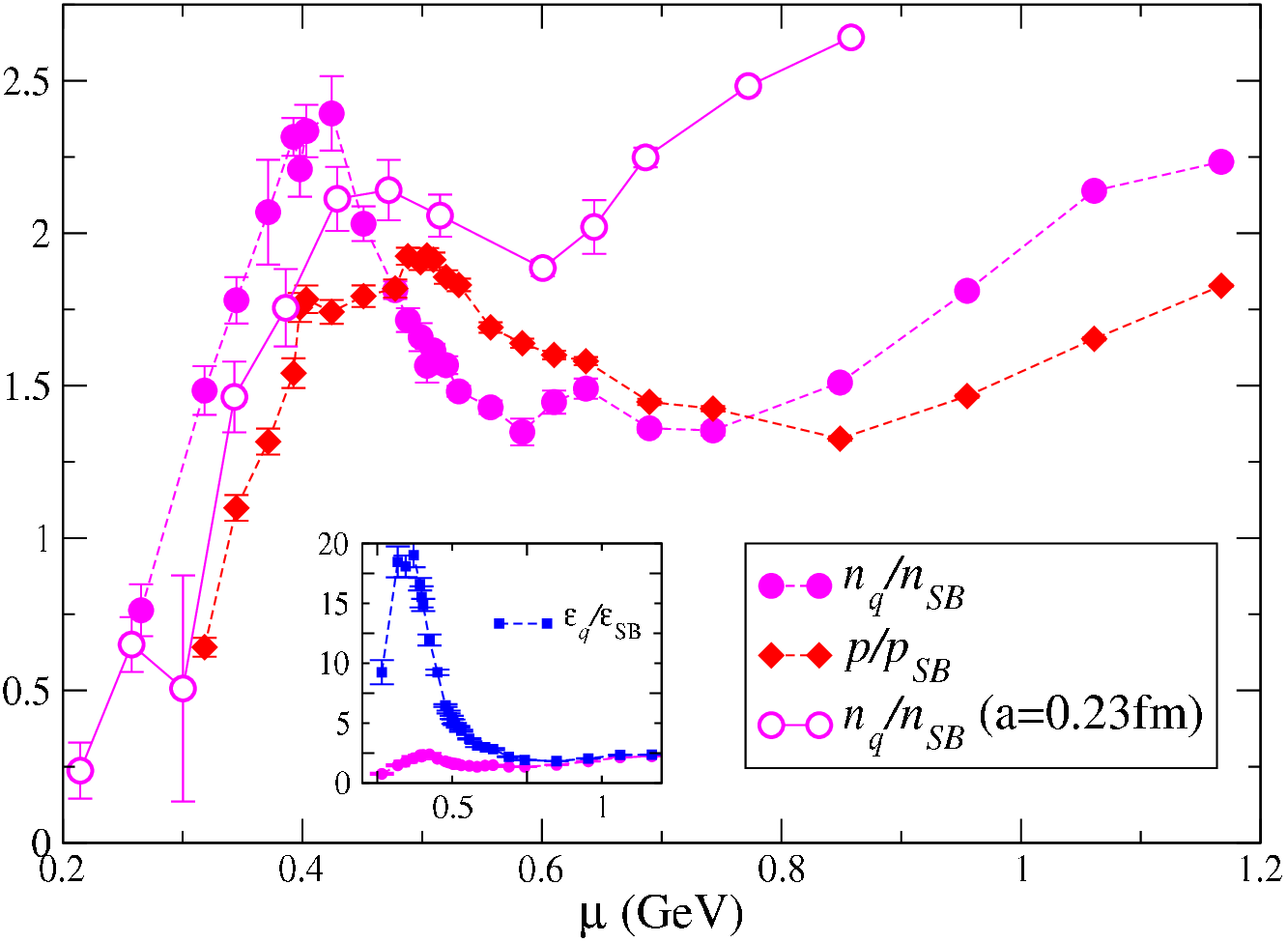}
    \caption{(Color online) $n_q/n_{SB}$ and $p/p_{SB}$ vs. $\mu$ for QC$_2$D. Inset shows
$\varepsilon_q/\varepsilon_{SB}$ for comparison.}
   \label{gr:nice}
\end{figure}
Fig.~\ref{gr:nice} shows results for quark density and pressure as functions of
$\mu$, plotted as
ratios of the same quantities evaluated for free massless quarks on the same
lattice~\cite{Hands:2006ve}. In the $j\to0$ limit the onset is expected at 
$\mu_o a=0.34$ corresponding to $\mu_o\simeq360$MeV; the observation of $n_q,p>0$ 
for $\mu<\mu_o$ is an artifact of working with $j\not=0$. Beyond onset 
the ratio $n_q/n_{SB}$ rises to a peak at $\mu\approx400$MeV, 
then falls to a
plateau beginning at $\mu_Q\approx530$MeV, which continues until $\mu_D
\approx850$MeV where it starts to rise again. If following the arguments
presented above
we associate the plateau with the setting in of degenerate
matter then we identify a BEC/BCS crossover at
$\mu_Q\approx530$MeV, corresponding to a quark density
$n_q\approx4 - 5\mbox{fm}^{-3}$, ie. roughly 10 times nuclear density.
In order to specify these numbers with greater precision, as well as the
statistical errors manifest in the error bars of Fig.~\ref{gr:nice} we need to
establish control over lattice artifacts by taking the continuum limit $a\to0$.
So far we have data from only two lattice spacings; results for $n_q/n_{SB}$
taken with $a=0.23$fm plotted in Fig.~\ref{gr:nice} show that there is
reasonable scaling for $\mu\lapprox400$MeV -- indeed, the small difference in
the physical $j$ between the two ensembles has maximal impact for
$\mu\approx\mu_o$~\cite{Kogut:2000ek,Hands:2006ve}. The situation at larger $\mu$ will be
discussed further below.

In contrast to $\chi$PT,  the quark contribution to the energy density
$\varepsilon_q$ exceeds the
free field value by almost a factor of 20 for $\mu\gapprox\mu_o$, as shown in
the inset of 
Fig.~\ref{gr:nice}; it should be remarked here that unlike $n_q$ and $p$,
$\varepsilon$ is subject to a multiplicative quantum correction known as a Karsch coefficient
\footnote{there is also an additive correction which has been subtracted by
requiring $\varepsilon(\mu=0)=0$.}
which is still to be calculated for this system, though its renormalised 
value is unlikely to
differ by more than 50\%. In any case, since the Karsch coefficient is
$\mu$-independent, the shape of the curve will remain the same. Because of this
unexpected behaviour at small $\mu$, the energy per
quark $\varepsilon_q/n_q$ exhibits a shallow but robust minimum for $\mu>\mu_Q$, 
a feature completely absent in the model governed by Eqns.
(\ref{eq:xPT1}-\ref{eq:SB1}).

\begin{figure}[tb]
    \centering
    \includegraphics[width=7.8cm]{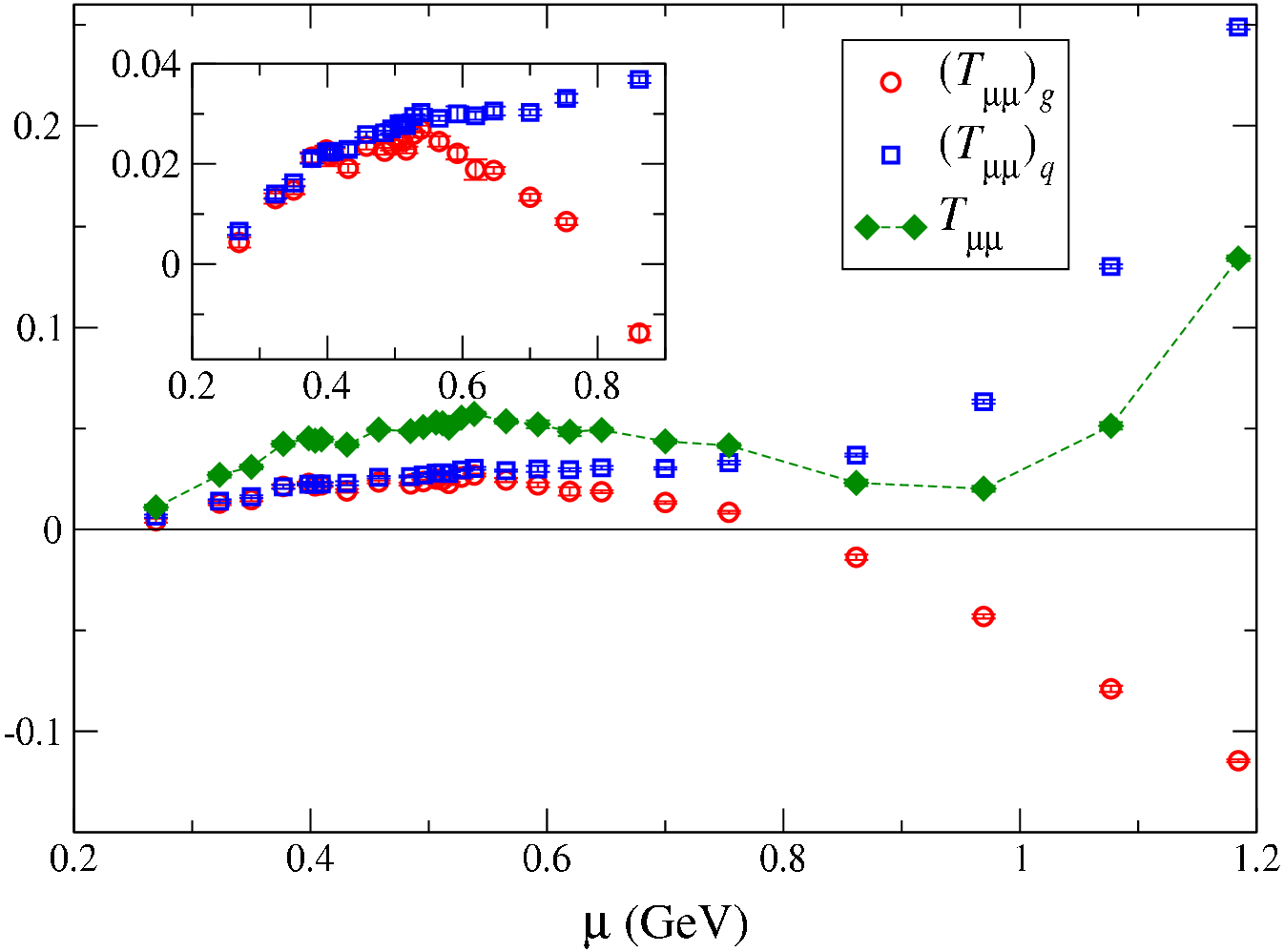}
\caption{(Color online) Conformal anomaly $T_{\mu\mu}a^4$ vs.
$\mu$, plotted together with separate quark and gluon contributions.}
\label{gr:conformal}
\end{figure}
Fig.~\ref{gr:conformal} plots 
the conformal
anomaly $T_{\mu\mu}=(T_{\mu\mu})_g+(T_{\mu\mu})_q$, with
\begin{eqnarray}
(T_{\mu\mu})_g&=&
-a{{\partial\beta}\over{\partial a}}\biggr\vert_{LCP}\times
{{3\beta}\over N_c}\mbox{Tr}\langle\Box_t+\Box_s\rangle;\\
(T_{\mu\mu})_q&=&
a{{\partial\kappa}\over{\partial a}}\biggr\vert_{LCP}\times
\kappa^{-1}(4N_fN_c-\langle\bar\psi\psi\rangle),
\end{eqnarray}
where $LCP$ denotes the beta-function is determined along a line of constant
physics, and once again a vacuum contribution must be subtracted.
The required beta-functions are estimated from the parameters
of the two matched lattices to be -0.85(17) ($g$) and 0.042(9) ($q$)
(these errors are not included in Fig.~\ref{gr:conformal}).

Fig.~\ref{gr:conformal} shows that for $\mu\lapprox\mu_Q$ the contribution of both quarks
and gluons to $T_{\mu\mu}$ is initially positive and increasing with $\mu$; 
here both quarks and gluons are contained within tightly-bound
non-relativistic bosons, implying that $\varepsilon>3p$.
For $\mu\gapprox\mu_Q$, however,
their behaviour diverges sharply (see inset),
which could possibly be explained by quark and gluon degrees of freedom now
being governed by differing quantum statistics.
In fact, 
the gluon data are very well approximated over the whole $\mu$-range by a parabola, and $(T_{\mu\mu})_g$
accordingly becomes negative for $\mu\gapprox\mu_D$. 
This change in sign  has also been predicted using $\chi$PT and the property of asymptotic
freedom~\cite{Metlitski:2005db}
(another way of understanding the necessity for $(T_{\mu\mu})_g$ to change sign
is that the plaquette must revert to its quenched value in the limit $\mu\to\infty$
where quantum corrections due to quarks are Pauli-blocked~\cite{Hands:2006ve}).
At the same point there is 
a very sharp change in the behaviour of $(T_{\mu\mu})_q$, which had been
approximately constant for $\mu_Q\lapprox\mu\lapprox\mu_D$.
Fig.~\ref{gr:conformal} shows
that at large $\mu$ the quark contribution dominates, so that $\lim_{\mu\to\infty}T_{\mu\mu}>0$.
This behaviour is not predicted by $\chi$PT (eg. Eqn.~(\ref{eq:xPT4}) and
Ref.~\cite{Metlitski:2005db}), although the positivity of $T_{\mu\mu}$ in this limit
is consistent with three-loop perturbation theory~\cite{Kurkela:2009gj}.

\begin{figure}[ht]
    \centering
    \includegraphics[width=8.2cm]{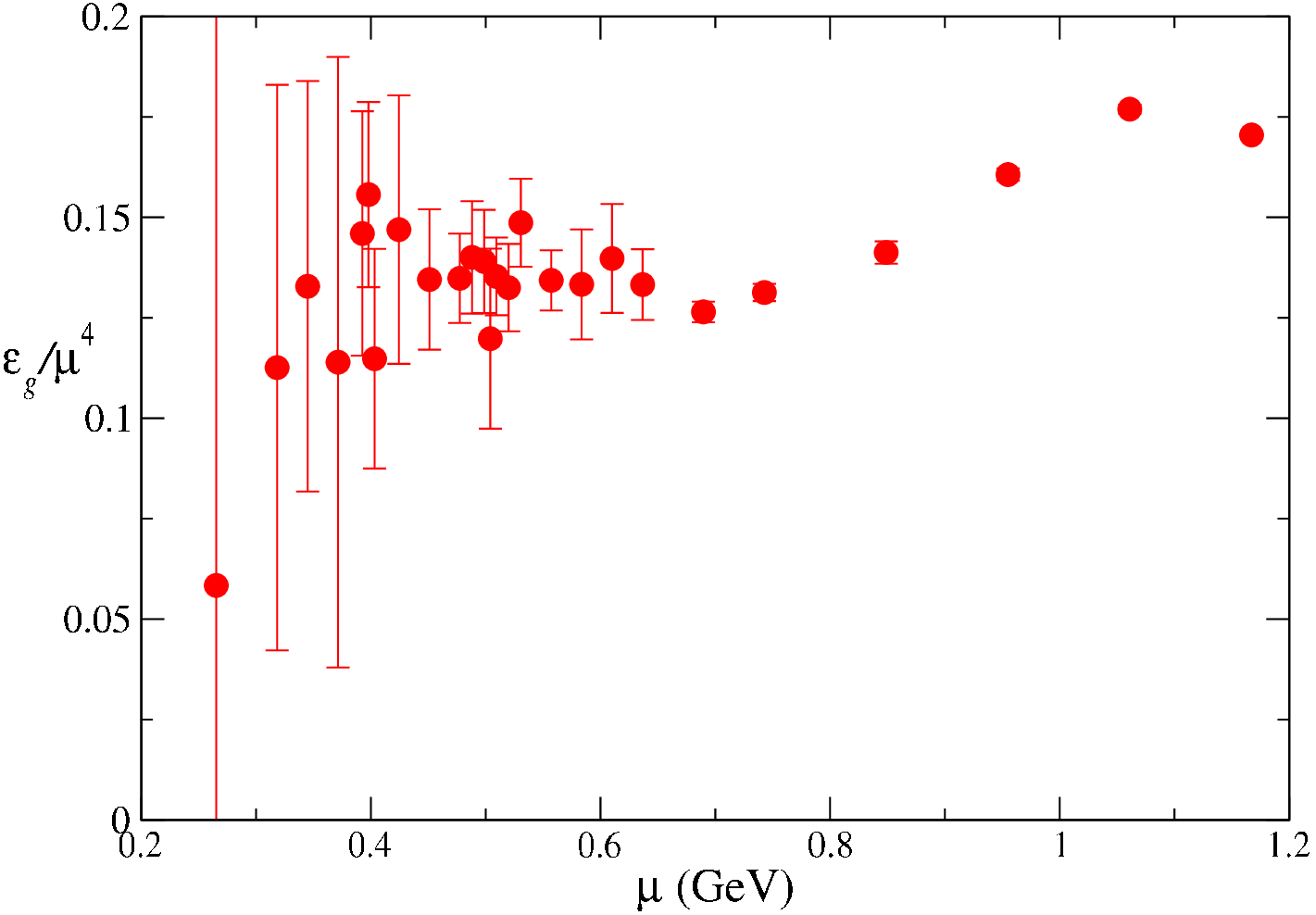}
    \caption{(Color online) Gluon energy density $\varepsilon_g/\mu^4$
versus $\mu$.}
   \label{gr:gluoned}
\end{figure}
Next consider the gluonic energy density given by
\begin{equation}
\varepsilon_g={{3\beta}\over N_c}\mbox{Tr}\langle\Box_t-\Box_s\rangle
\end{equation}
(once again, a $\mu$-independent Karsch coefficient is still to be determined).
Fig.~\ref{gr:gluoned} plots the dimensionless combination $\varepsilon_g/\mu^4$
against $\mu$; of course this quantity is not predicted either
in $\chi$PT or the free
quark gas. 
While we therefore have no quantitative theory of the gluonic contribution to QC$_2$D
thermodynamics at $\mu\not=0$, we would expect its relative importance to
increase across a deconfining transition.
In fact, the ratio is remarkably constant over a wide range of
$\mu$, consistent with dimensional analysis; 
in particular there is no sign of singular behaviour at
$\mu=\mu_Q$, although there is a systematic rise for 
$\mu\gapprox\mu_D$.

\begin{figure}[ht]
    \centering
    \includegraphics[width=8.0cm]{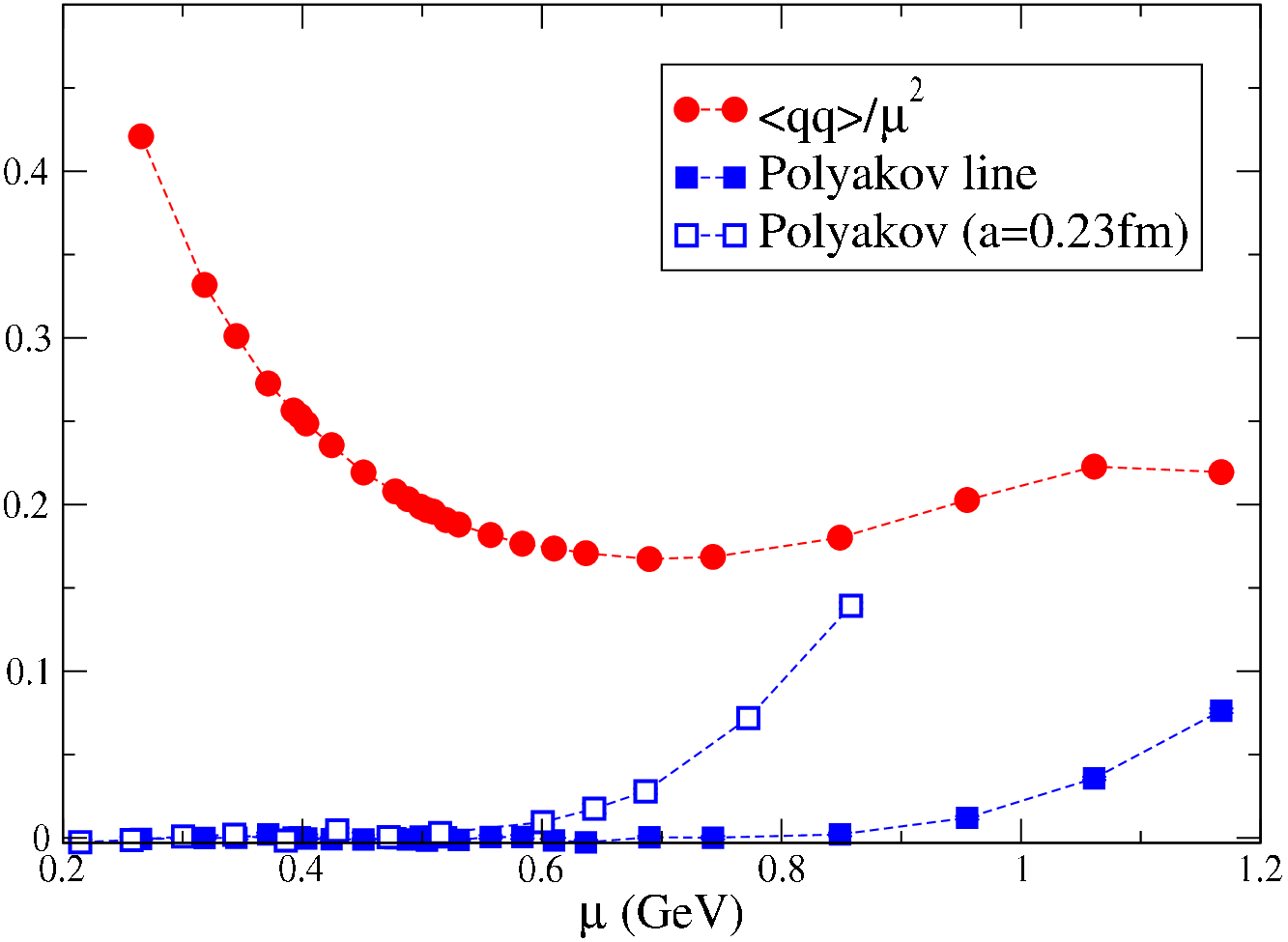}
    \caption{(Color online) Superfluid order parameter $\langle qq\rangle/\mu^2$ and Polyakov
line versus $\mu$.}
   \label{gr:orderps}
\end{figure}
Fig.~\ref{gr:orderps} plots quantities giving information on the
nature and symmetries of the ground state. In the limit $j\to0$, the diquark
condensate $\langle qq\rangle$ is an order parameter for the spontaneous
breaking of U(1)$_B$ leading to baryon number
superfluidity. Although the data of Fig.~\ref{gr:orderps} are taken with
$j\not=0$, implying some care must be taken with the extrapolation $j\to0$ at
small $\mu$~\cite{Hands:2006ve}, we are confident that this symmetry is broken
for all $\mu>\mu_o$. The approximate flatness of the curve for
$\mu_Q\lapprox\mu\lapprox\mu_D$ is then evidence for a scaling $\langle
qq\rangle\propto\mu^2$ similar to Eqn.(\ref{eq:BCS}). We take this as an indication
that in this region the system consists of degenerate quark matter with a Fermi
surface disrupted by a BCS instability.

The Polyakov line is an order parameter for deconfinement in the limit of
infinitely massive quarks  -- away from this limit it continues to yield
information on the free energy of an isolated color source.
Fig.~\ref{gr:orderps} shows that QC$_2$D remains confined for $\mu<\mu_D$, but
that there appears to be a transition to a deconfined state for chemical
potentials in excess of this value. In physical units $\mu_D\approx850$MeV,
corresponding to quark density $n_q\approx16$ -- 32fm$^{-3}$, some 35 -- 70 times
nuclear density.

To summarise, the simulations suggest that QC$_2$D has three distinct
transitions (or at least crossovers). The first, at $\mu=\mu_o$, is
a firmly established second order
phase transition (in the limit $j\to0$) from vacuum to a BEC superfluid, and is
described accurately for the most part by $\chi$PT (the quark energy density
$\varepsilon_q$ looks to be an important exception). Since the pion is not
especially light with our choice of lattice parameters, implying only a moderate  separation of
Goldstone and hadronic mass scales, the $\mu$-window within which
the BEC is favoured is not particularly wide.

The second transition
at $\mu=\mu_Q$ looks like a BEC/BCS
crossover to form a ground state where the scalings of the observables $n_q(\mu)$,
$p(\mu)$, $\varepsilon_q(\mu)$ and $\langle qq(\mu)\rangle$ all suggest it is
formed of degenerate quark matter with a well-defined Fermi sphere, albeit one
whose surface is disrupted by a BCS condensate. 
We note that effective treatments based on both $\chi$PT and NJL models
predict $n_q/n_{SB}$ to be monotonic decreasing in this regime, and are unable
to fit the lattice data~\cite{Andersen:2010vu}.
The distinct nature of this region is also supported by the diverging behaviours
of $(T_{\mu\mu})_g$ and $(T_{\mu\mu})_q$ for $\mu\gapprox\mu_Q$ seen in
Fig.~\ref{gr:conformal}, although the reason for the peculiar
behaviour of $(T_{\mu\mu})_q$ is not well understood at present.  The
transition at $\mu\approx\mu_Q$ is most likely a smooth crossover, but
the exact nature of this putative transition requires further study,
in particular a careful extrapolation to the limit of zero diquark
source.

The third transition at $\mu=\mu_D$ is
signalled by a change in the scaling of the thermodynamic observables, notably $\varepsilon_g(\mu)$ and
$(T_{\mu\mu})_q$, a change in the sign of $(T_{\mu\mu})_g$,
and a non-zero Polyakov
loop. For $\mu>\mu_D$ the system consists of deconfined quark matter. 

An immediate concern is the validity of the deconfining transition in the
continuum limit. Fig.~\ref{gr:orderps} shows that with $a=0.23$fm
$\mu_D\approx600$MeV, and is practically indistinguishable from $\mu_Q$; for
this reason 
only a deconfined quark matter phase was identified in Ref.~\cite{Hands:2006ve}
(Cf. Fig.~\ref{gr:nice}).
In both cases, however, the quark density in lattice units $n_qa^3=0.17$ (coarse)
or 0.20 (fine), well short of the value $2N_cN_f$ signifying that lattice
saturation artifacts have set in -- indeed for $a=0.186$fm trends in all
observables look smooth out to $\mu a=1.0$ corresponding to $\mu=1.06$GeV. 
It is therefore plausible that the
observed difference in $\mu_D$ is physical, and due to the differing
temperatures of the two lattices used.

Between $\mu_Q$ and $\mu_D$ the system
resembles the quarkyonic matter recently postulated on the basis of
large-$N_c$ arguments~\cite{McLerran:2007qj}; 
namely a state of degenerate matter which is 
also confined, so that
excitations above the ground state remain color singlet. 
Because we have used Wilson fermions (with no manifest chiral symmetry)
we are unfortunately unable at this point to test whether chiral symmetry is 
restored, another important aspect of the quarkyonic
hypothesis; we note however that even in a conventional scenario $\chi$PT
predicts
$\langle\bar\psi\psi\rangle\propto\mu^{-2}$  for
$\mu\geq\mu_o$~\cite{Kogut:2000ek}, which is likely
to be difficult to distinguish from true chiral symmetry restoration in this
region,
particularly if the transition is a crossover.

The apparent
sensitivity of the value of $\mu_D$ to small changes in temperature is
consistent with the very weak curvature of the phase boundary between confined
and deconfined phases postulated in that work, and observed in
a recent study of QC$_2$D
matter using the PNJL model~\cite{Kenji:2009}.

An interesting issue is whether QC$_2$D is special in that the $N_c$-quark bound
states required by color confinement are also favoured by the more general
renormalisation group argument that 2-body interactions are the only relevant
ones close to a Fermi surface~\cite{Shankar:1993pf}. QC$_2$D is also exceptional,
of course, because since $\langle qq\rangle$ is gauge singlet there is no CSC phase.
Whatever the outcome, 
to our mind the study of deconfinement in this hitherto-unexplored physical
regime promises to be fascinating.

\begin{acknowledgments}
This project was enabled with the assistance of IBM Deep Computing.
S.K. was supported by the National Research Foundation of Korea grant funded
by the Korea government (MEST) No. 2009-0074027.
We benefitted greatly from discussions with Ernst-Michael Ilgenfritz.
\end{acknowledgments}

\end{document}